# Inter-communication between Programmable Logic Controllers using IoT technologies: A Modbus RTU/MQTT Approach

Marios Tyrovolas1 and Tibor Hajnal2

*Abstract—* Internet of Things (IoT) can be widely used in various applications such as manufacturing industry, achieving high operational efficiency and increased productivity. The exploitation of IoT paradigm made more feasible the use of distributed control systems (DCSs) where more than one PLCs implement an industrial application. In case of having more than one PLC, it is obligatory to ensure their inter-communication. In this paper, a solution for PLCs' communication using Message Queuing Telemetry Transport (MQTT) IoT-protocol is presented. An experimental industrial plant was used, controlled by two PLCs. For the process' implementation, two Siemens gateways Simatic IoT2020 were integrated and communicated with the PLCs through Modbus Remote Terminal Unit (RTU) industrial protocol. Finally, data exchange between the two gateways was established using MQTT protocol.

## I. INTRODUCTION

Nowadays, more people are taking advantage of modern communication networks' capabilities using "smart" devices in various sectors of society. This continuous upgrowth, allows the introduction of new working, communication and entertainment methods leading to new lifestyle, while at the same time expands the Internet's boundaries resulting in the transition from "Internet of Humans" to "Internet of Things" (IoT) [1]. IoT refers to diverse devices interconnection in order to exchange data in real time without human intervention [2]. The creation of such a network can play a very important role in industries. In particular, the Industrial Internet of Things (IIoT) can significantly improve industrial devices' connectivity, process scalability and overall factory efficiency. Despite the advantages offered from IIoT technologies, industries still rely on their legacy systems consisting of Programmable Logic Controllers (PLCs). The term PLC is referred to a microcomputer used to control an industrial plant, offering increased quality and reliability leading to production cost reduction [3]. Consequently, they are considered as an integral part of any industrial environment and cannot be absent from any future IIoT solution.

The advent of IIoT, allows communication between industrial devices leading to a new trend for increased use of DCSs at machine's level. Specifically in DCSs, instead of using one PLC to control an automation system consisting of multiple actuators and robotic manipulators, more PLCs are used to control the plant. In such systems, PLCs must acquire the ability to interact with each other according to external events, such as last-time changes in customer's orders and therefore improving system's response. Therefore, it is necessary to solve PLCs' inter-communication problem.

In order to enable communication between PLCs, special developed fieldbus protocols (EtherCat, Profinet, Modbus, EtherNet/IP etc.) providing real-time communication with delays of several microseconds have been proposed [4] – [8]. By considering an IoT-based direction for future industries, IoT protocols should be utilized for independent units' inter-connection in a manufacturing process. At the same time, the use of these protocols could be extended beyond the factory boundaries by giving the opportunity to the user to remotely control the automation system and monitor the proper operation of production. This could lead to the creation of a network where both industrial devices and users can communicate with each other through a unique protocol. This paper proposes the use of MQTT IoT-protocol on an experimental industrial plant and emphasizes in the PLCs' inter-communication. The paper is organized as follows: Section II presents the system layout while Section III describes the technical details of its implementation. Finally, Section IV comments on the conclusions reached during the implementation.

## II. SYSTEM LAYOUT

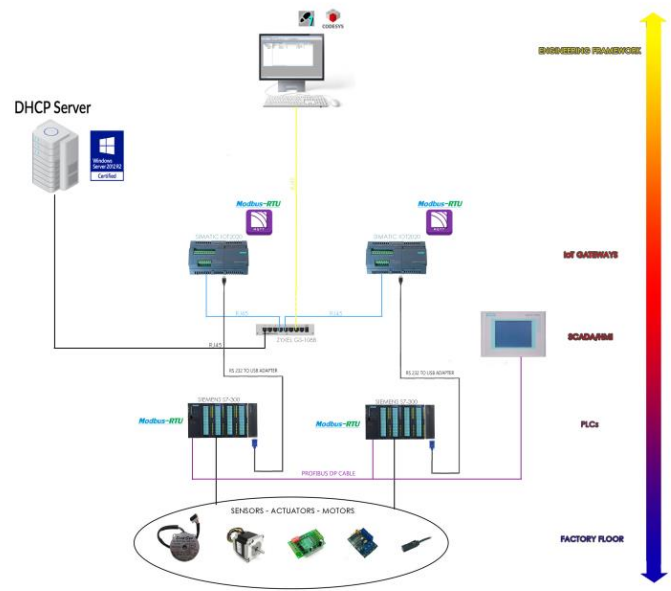

Figure 1. High-level overview of the industrial plant

The experimental automation system consists of 7 sub-units performing either rotational or linear motion for object transportation. As shown in Fig. 2, all transport movements are performed using twelve stepper motors which in turn are controlled by two PLCs. The first five motors are controlled by the first PLC and the remaining seven motors by the second one. In order to implement this application, communication between the two controllers had to be established. For this reason, two Siemens gateways (Simatic IoT2020) were used for data exchange (Fig.1). Simatic IoT2020 collects data through its various communication ports from PLCs located in this case-study [9]. This can be achieved by using secure industrial protocols such as Modbus, Profinet etc.

[1]Marios Tyrovolas, is with Electrical and Computer Engineering Department, University of Patras, GR-26504, Patras, Greece
[2]Tibor Hajnal is with Control & Automation Group, Saudi Aramco, Jeddah, Kingdom of Saudi Arabia.
Corresponding Author's Email: `up1020654@upnet.gr`

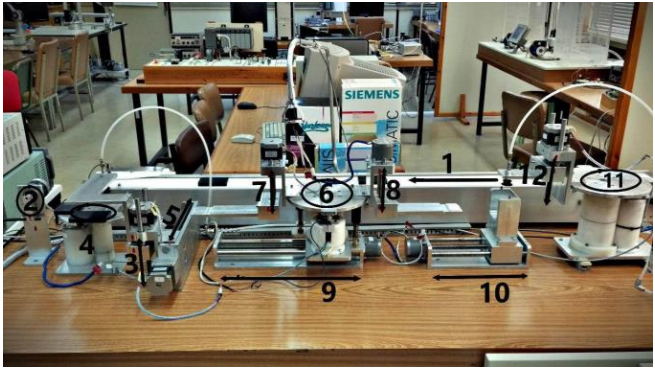

Figure 2. Deliberated Industrial Plant consisting of 12 motors

## III. IMPLEMENTATION DETAILS

The idea for system's operation be as follows: When the first PLC completes the movement of its individual units, a flag is activated ("HOLDING REGISTERS".HREG_40001) indicating the accomplishment of the process. The change of flag's value is notified to the second controller, which, in turn, executes its own sequence of actions. When motor 9 or 11 completes its operation, the corresponding flags of the second PLC are activated and by informing their value to the first controller, the procedure is repeated.

An arising question is "How the flags' value change is being known to the controllers?". The data exchange between PLCs is achieved through two gateways which communicate utilizing IoT technologies. However, in order to be able to read the new value of the flags but also to inform the controllers about this change, communication between Simatic IoT2020 and PLCs was enabled using serial communication protocol Modbus RTU.

### A. Analysis of Modbus RTU protocol

Modbus RTU is an open industrial serial communication protocol based on master-slave architecture. According to this architecture, slaves never transmit data unless they first receive a request from the master who has complete control over the data flow [10]. The query message towards the slave consists of [11]:

- Device Address: The slave's address takes values from 1 to 247 (decimal form) letting the master know which slave is responding.

- Function Code: Indicates how the master interacts with slave.

- Data: This field contains any additional information that the slave will need to perform the function.

- Error Check: Provides a method for validating the content of the message. Master calculates the checksum and sends it to the slave, which, in turn, recalculates it and compares the two values. If any difference between the two values is detected, the slave will not respond to the master's request

### B. Implementation of Modbus RTU in PLC

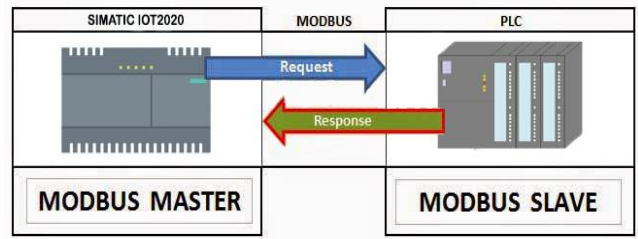

Figure 3. The master-slave and query-response relation between Modbus devices

In this case-study, Modbus RTU communication via RS-232 cable between Simatic IoT2020 (Modbus master) and PLC (Modbus Slave) was implemented. Simatic S7-300 became Modbus slave by using software driver and communication processor module CP-340. This driver consists of three function blocks:

- FB36 "MSL340" (Modbus Slave Communication Function Block)

- FB3 "P_SEND" (Siemens standard send function block for CP340)

- FB2 "P_RCV" (Siemens standard receive function block for CP340)

"MBSL340" is the only function block which called by the user's program. The "P_SEND" and "P_RCV" function blocks are called within "MBSL340" [12].

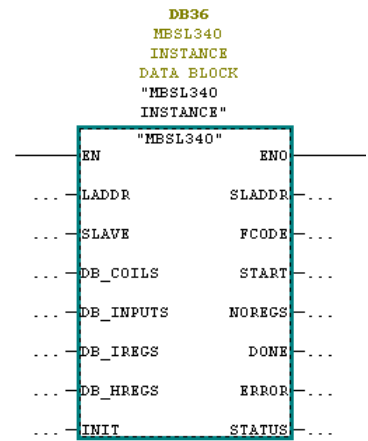

Figure 4. "MBSL340" Function Block

Therefore, to establish PLC as Modbus slave, "MBSL340" was called in the beginning of Organization Block 1 (OB1) and protocol's general parameters (Baud Rate, Parity, Data Bits, Stop Bits) were configured properly. These parameters must fit with the relevant parameters of the Master as any mismatch results in communication failure. Baud Rate was set at 9600 bps while at the same time even parity was selected. Furthermore, the character field consists of 8 Data Bits and 1 Stop Bit.

## C. Implementation of Modbus RTU in Simatic IoT2020

To set Simatic IoT2020 as Modbus Master, Codesys platform was used. Codesys is a development environment for programming controller applications according to the international industrial standard IEC 61131-3 where integrated compilers transform the user's application code into native machine code (binary code) [13]. For the proposed implementation, Modbus interface was inserted in the project and subsequently both protocol's general parameters and the two communication endpoints were configured. As it is illustrated to Fig.5 and Fig.6, response time, time between frames and slave address were configured.

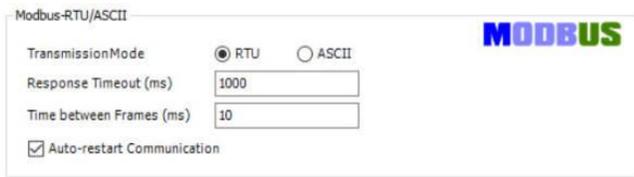

Figure 5. Settings for Modbus Master in Codesys

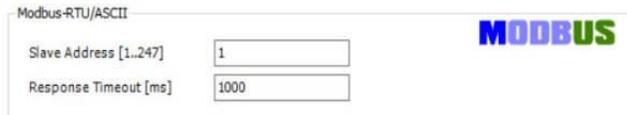

Figure 6. Settings for Modbus Slave in Codesys

Furthermore, two communication channels between master and slave for reading or writing data to PLCs were defined. The gateways read the flags' value by sending the function code 03 (Read Holding Registers) while for the updated value's entrance, the first gateway sends the function code 15 (Write Multiple Coils) and the second one sends the code 05 (Write Single Coil). Function code 15 was chosen as the first controller awaits two flag values (if motor 9 or 11 returned to their reference point) while function code 05 was used as the second PLC expects one flag value (if the first controller completed the movement of its individual units).

## D. Analysis of MQTT protocol

For data exchange between the two controllers, gateways' intercommunication had to be established. This was achieved by using IoT technologies and specifically MQTT communication protocol. MQTT is a lightweight and "compact" messaging protocol for transferring data remotely when low bandwidth is required. It focuses on minimum overhead (2 bytes header) and reliable communication between low-level embedded devices thus leading to efficient utilization of network resources [14]. It is based on publish-subscribe architecture where clients communicate with each other through a server called MQTT Broker. A client can either send data (publisher) or receive data (subscriber) which are organized into a hierarchy of topics. When publisher has new data, it sends a message to the Broker which in turn, distributes them to any subscriber of the specific topic so the two endpoints are completely decoupled [15].

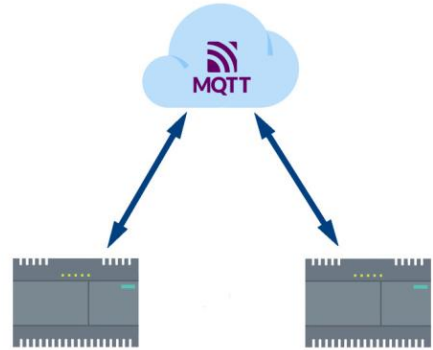

Figure 7. Gateways' communication via MQTT

## E. Implementation of MQTT in Simatic IoT2020

MQTT was also implemented using Codesys platform. From the user's program only one function block is called which is programmed according to IEC 61131-3 standard.

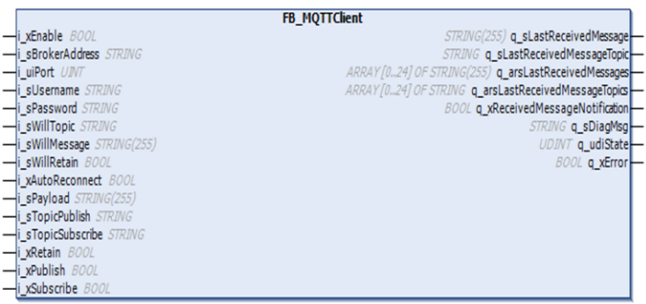

Figure 8. MQTT Client Function Block

After the required function block (FB_MQTTClient) was inserted into the project, its input parameters were defined. As shown in Fig. 7, the xEnable parameter for the manual connection between gateway and MQTT Broker was set along with MQTT Broker's IP address (i_xBrokerAddress) and MQTT's port (1883). Next, two boolean variables (xPublish and xSubscribe) for publishing and subscribing messages were set, while at the same time publish and subscribe topics for each gateway were defined. Furthermore, i_xRetain parameter was defined permanently to state 0 (false) for the system's proper operation and message's content (payload) was set as the value of the flags read from controllers. Specifically, the first gateway reads Holding Register 4001 while the second gateway reads both Holding Registers 4001 and 4002. Therefore, in the program of the first Simatic IoT2020, an input (Register4001) with address %IW0 was set while at the same time, two inputs (Register4001 and Register4002) with addresses %IW0 and %IW1 respectively were set in the program of the second IoT2020. In addition, every gateway publishes data only when a change in their value occurs. The proposed technology was experimentally verified resulting in the proper operation of the industrial plant without having any problems. Specifically, the motors operated correctly while at the same time losses in micro-motion accuracy of the units didn't appeared.

## IV. CONCLUSIONS

In this study, interconnection between two PLCs for controlling an experimental industrial plant by using Modbus RTU and MQTT protocols was proposed. As it was observed, MQTT can be used in remote control applications, however, it doesn't have a built-in mechanism for updating the variables of a PLC. Therefore, it needs a tool that will receive the published message, process it and access the PLC to update the variables.

For a better comparison, it would be advisable to focus on interconnecting Programmable Logic Controllers and other industrial devices through more compatible industrial protocols such as OPC UA. Specifically, we could compare data exchange time between PLCs and therefore system's response time. This will give users more flexibility in deploying the applications that will meet the needs of each automation system.